\documentclass{PoS}

\title{High Precision Statistical Landau Gauge Lattice Gluon
   Propagator Computation}

\ShortTitle{High Precision Statistical Landau Gauge Lattice Gluon
   Propagator Computation}

\author{David Dudal \\
        KU Leuven Kulak, Department of Physics, Etienne Sabbelaan 53 bus 7657, \\8500 Kortrijk, Belgium \\
        Ghent University, Department of Physics and Astronomy, Krijgslaan 281-S9, \\9000 Gent, Belgium\\
        E-mail: \email{david.dudal@kuleuven.be}}

\author{Orlando Oliveira, \speaker{Paulo J. Silva}\\
       CFisUC, Departamento de F\'{i}sica, Universidade de Coimbra, 3004-516 Coimbra, Portugal\\        
        E-mail: \email{orlando@uc.pt}, \email{psilva@uc.pt}}

\abstract{We report on results for the Landau gauge gluon propagator computed from large statistical ensembles and look at the compatibility of the results with the Gribov-Zwanziger tree level prediction for its refined and very refined versions. Our results show that the data is well described by the tree level estimate only up to momenta $p \lesssim 1$ GeV, while clearly favouring the so-called Refined Gribov-Zwanziger scenario. We also provide a global fit of the lattice data which interpolates between the above scenario at low momenta and the usual continuum one-loop renormalisation improved perturbation theory after introducing an infrared log-regularizing term.}

\FullConference{The 36th Annual International Symposium on Lattice Field Theory - LATTICE2018\\
		22-28 July, 2018\\
		Michigan State University, East Lansing, Michigan, USA.}

\begin{document}

\section{Introduction and Motivation}

The Landau gauge gluon propagator has been intensively studied using lattice simulations in the past recent years. The picture emerging being a propagator that at small
momenta is finite, therefore suppressed relative to the perturbative calculation, and non-vanishing at zero momentum. These results can be understood as due
to the dynamical generation of mass scales, that can also be interpreted as a running gluon mass, that regulate  the would-be infrared 
singularities~\cite{Bogolubsky:2009dc,Dudal:2010tf,Oliveira:2010xc,Oliveira:2012eh,Cucchieri:2011ig,Siringo:2015wtx,Duarte:2016iko,Chaichian:2018cyv,Dudal:2018cli}.

The interpretation of the lattice gluon propagator needs a proper non-perturbative quantisation of the Yang-Mills theories, a problem not yet completely solved
due to the presence of the so-called Gribov copies~\cite{Gribov:1977wm}. An improvement over the standard construction of the Green's function generating functional,
i.e. the Faddeev-Popov trick~\cite{Faddeev:1967fc}, that resulted in a local renormalized action was suggested in~\cite{Zwanziger:1989mf} and lead to the construction
of a family of actions named generally Gribov-Zwanziger actions; for a review see e.g.~\cite{Vandersickel:2012tz} and references therein. These actions introduce
mass scales that regulate the Yang-Mills theory at low energy. Indeed, as discussed 
in~\cite{Dudal:2010tf,Cucchieri:2011ig,Dudal:2018cli}, the Refined Gribov-Zwanziger action and the Very Refined Gribov-Zwanziger action~\cite{Dudal:2008sp,Capri:2017bfd}
 are compatible with lattice simulations
if not over the full range of momenta, at least for momenta up to $\sim 1$ GeV, i.e. for the two point correlation function they provide analytical results that are compatible with the 
low energy behaviour of the gluon propagator observed in lattice simulations. 

The initial studies of the lattice simulations versus Gribov-Zwanziger action had a relatively small
statistics, typically the number of configurations per ensemble being of the $\mathcal{O}$(100) or smaller. The question we would like to address here being if the early
studies~\cite{Dudal:2010tf,Cucchieri:2011ig} results are still valid when one increases significantly the number of configurations per ensemble. In order to try to answer it,
we consider two large physical volume lattice simulations performed with $\beta = 6.0$, that corresponds to a lattice spacing of $a = 0.1016(25)$ fm ($1/a = 1.943$ GeV) 
measured from the string tension, and for $64^4$ and $80^4$ lattices, whose physical size being $L = 6.57$ fm and $L = 8.21$ fm, respectively. For the simulation using the
smaller physical volume we consider an ensemble with 2000 gauge configurations rotated to the Landau gauge, while for the largest physical volume the ensemble has 550 gauge configuration rotated to the Landau gauge. Herein, we resume the results reported in~\cite{Dudal:2018cli}, where the reader can find
further details on the calculation.
Besides the investigation of the compatibility of the Gribov-Zwanziger functional form with the high precision lattice data, we also address the question of extending the
predicted functional forms to cover the full range of momenta accessed in lattice simulations, while keeping the right perturbative tail in the ultraviolet regime.

In the Landau gauge, the gluon propagator is defined as
\begin{equation}
  \langle A^a_\mu (\hat{p}^\prime) ~ A^b_\nu (\hat{p})  \rangle = V ~ \delta^{ab} ~ \delta( \hat{p}^\prime + \hat{p} ) ~ P_{\mu\nu} (p) ~ D(p^2) \ ,
\end{equation}  
where $\hat{p}_\mu =  (2 \, \pi / a L ) \, n_\mu$ with $n_\mu = 0, \cdots, L-1$ is the lattice momentum, $p = (2/a) \, \sin ( \pi \, n_\mu / L )$ is the so-called
continuum momentum, $P_{\mu\nu}(p)$ is the orthogonal projector and $V = L^4$ is the lattice volume. All the lattice data reported is renormalised in the MOM-scheme, where
$    \left. D(p^2)  \right|_{p^2 = \mu^2} = 1 / \mu^2 \ , $ and we have used $\mu = 3$ GeV as renormalisation scale.

\section{Refined Gribov-Zwanziger, Very Refined Gribov-Zwanziger and Lattice Data}

The tree level prediction of the Refined Gribov-Zwanziger for the gluon propagator being
\begin{equation}
   D(p^2) = Z \, \frac{ p^2 + M^2_1}{p^4 + M^2_2 ~ p^2 + M^4_3}
   \label{EQ:RGZ}
\end{equation}
and, in order to study the compatibility  of this expression with the lattice data, we perform a correlated fit taking into account the lattice data for
momenta in $[ 0 \, , \, p_{max} ]$ and change $p_{max}$ monitoring the corresponding value of the $\chi^2/d.o.f.$ Furthermore, in the fits
we consider two cases, where $Z$ is left as a free parameter and where we set $Z = 1$. The fits for a $\chi^2/d.o.f. \lesssim 2$ can
be seen on Tab.~\ref{tab:rgzfits}. For the simulation using the smaller physical volume one can claim a
\begin{displaymath}
\begin{array}{l@{\hspace{0.5cm}}l@{\hspace{0.5cm}}l@{\hspace{0.5cm}}l@{\hspace{.9cm}}l}
  Z = 1.088(58) & M^2_1 = 2.16(22)    & M^2_2 = 0.478(21)  & M^4_3 = 0.261(13) & p_{max} = 1.00  \\
  Z = 1               & M^2_1 = 2.521(28) & M^2_2 = 0.5082(90) & M^4_3 = 0.2795(27) & p_{max} = 1.00
\end{array}  
\end{displaymath}
where all quantities are given in powers of GeV. For the simulation using the larger physical volume one gets
\begin{displaymath}
\begin{array}{l@{\hspace{0.5cm}}l@{\hspace{0.5cm}}l@{\hspace{0.5cm}}l@{\hspace{.9cm}}l}
  Z = 0.957(66) & M^2_1 = 2.73(34)     & M^2_2 = 0.527(29)   & M^4_3 = 0.290(16) & p_{max} = 1.10  \\
  Z = 1               & M^2_1 = 2.525(36) & M^2_2 = 0.510(11)  & M^4_3 = 0.2803(34) & p_{max} = 1.10
\end{array}  
\end{displaymath}
where all quantities are measured in powers of GeV. The fitted parameters are in good agreement for the two simulations,
with the case where $Z = 1$ producing closer results for the two physical volumes.
The weighted average of the fitted parameters results in
$Z = 1.027(44)$, $M^2_1 = 2.38(19)$ GeV$^2$, $M^2_2 = 0.499(17)$ GeV$^2$ and $M^4_3 = 0.274(10)$ GeV$^4$ or
$Z = 1$, $M^2_1 = 2.523(22)$ GeV$^2$, $M^2_2 = 0.5090(70)$ GeV$^2$, $M^4_3 = 0.2799(21)$ GeV$^4$.

We conclude confirming the results of~\cite{Dudal:2010tf,Cucchieri:2011ig} now for a high statistical calculation that the tree level
refined Gribov-Zwanziger prediction for the gluon propagator is compatible with the low energy lattice data up to momenta
$p \approx 1$ GeV.

The tree level prediction of the Very Refined Gribov-Zwanziger action for the gluon propagator reads
\begin{equation}
   D(p^2) =  \frac{ p^4 + M^2_1 ~ p^2 + M^4_2}{p^6 + M^2_5 ~ p^4 + M^4_4 ~p^2 + M^6_3} \ .
   \label{EQ:VRGZ}
\end{equation}
As discussed in~\cite{Dudal:2018cli}, we have observed that the fitting range where (\ref{EQ:VRGZ}) is compatible with the lattice data 
covers essentially the same range of momenta, i.e. $p \in [ 0 \, , \, 1]$ GeV. Furthermore, the fitted parameters are such that
$M^4_2 \approx 0$ GeV$^4$ and $M^6_3 \approx 0$ GeV$^6$ and (\ref{EQ:VRGZ}) reduces to (\ref{EQ:RGZ}) with the 
parameters $( M^2_1, \,  M^2_5 , \,  M^4_4)$ appearing in (\ref{EQ:VRGZ}) reproducing, within errors, the parameters
$( M^2_1, \,  M^2_2 , \,  M^4_3)$ that define (\ref{EQ:RGZ}). The reducing of (\ref{EQ:VRGZ}) to (\ref{EQ:RGZ}) suggests that the condensate 
named $\rho$ in~\cite{Dudal:2018cli} is real.

\begin{table}[t]
   \centering
   \begin{tabular}{l@{\hspace{0.75cm}} l@{\hspace{0.75cm}} l @{\hspace{0.5cm}} l @{\hspace{0.5cm}}l @{\hspace{0.5cm}} l @{\hspace{0.5cm}} l}
      \hline\hline
       $L$ & $p_{max}$  & $\nu$  & $Z$  & $M^2_1$  & $M^2_2$  & $M^4_3$ \\
       \hline
       64 & 0.50    & 1.84    &  $2.2  \pm 1.3$  &  $0.57  \pm  0.78$   &  $0.421 \pm 0.045$  & $0.14  \pm 0.11$ \\
       64 & 0.70    & 1.12    &  $1.50  \pm 0.17$  &  $1.19  \pm 0.28$    &  $0.417 \pm 0.027$  & $0.200  \pm 0.025$ \\
       64 & 0.80    & 1.14    &  $1.39 \pm 0.17$ &  $1.40   \pm 0.34$     &  $0.432 \pm 0.028$  & $0.216  \pm 0.025$ \\
       64 & 0.90    & 1.14    &  $1.199  \pm 0.084$ & $1.82  \pm 0.25$  &  $0.458 \pm 0.022$  & $0.243  \pm 0.015$  \\
       64 & 1.00    & 1.22    &  $1.088 \pm 0.058$ & $2.16 \pm 0.22$    &  $0.478 \pm 0.021$  & $0.261 \pm 0.013$  \\
       64 & 1.10    & 1.83    &  $0.959 \pm 0.062$ & $2.67 \pm 0.32$    &  $0.511 \pm 0.026$   & $0.285 \pm 0.015$  \\
       \hline
      80 & 0.50     & 0.45     & $2.69  \pm 0.35$   & $0.25\pm 0.13$    & $0.362 \pm 0.012$    & $0.077 \pm 0.028$ \\
      80 & 0.70     & 1.07     & $1.62 \pm 0.19$    & $1.03 \pm 0.28$    & $0.408 \pm 0.033$   & $0.186 \pm 0.029$ \\
      80 & 0.80     & 1.04     & $1.48 \pm 0.17$    & $1.25 \pm 0.30$    & $0.428 \pm 0.033$   & $0.206 \pm 0.027$ \\
      80 & 0.90     & 1.03     & $1.36 \pm 0.13$    & $1.48 \pm 0.29$    & $0.447 \pm 0.030$   & $0.224 \pm 0.022$ \\
      80 & 1.00     & 1.11     & $1.075 \pm 0.087$ & $2.26 \pm 0.34$   & $0.500 \pm 0.029$   & $0.269 \pm 0.018$ \\
      80 & 1.10     & 1.16     & $0.957 \pm 0.066$ & $2.73 \pm 0.34$   & $0.527 \pm 0.029$   & $0.290 \pm 0.016$ \\
      80 & 1.25     & 1.34     & $0.832 \pm 0.062$ & $3.42 \pm 0.44$   & $0.565 \pm 0.031$   & $0.315 \pm 0.016$ \\
      80 & 1.50     & 1.39     & $0.723 \pm 0.037$ & $4.29 \pm 0.37$   & $0.610 \pm 0.026$   & $0.341 \pm 0.012$ \\
      80 & 1.75     & 1.31     & $0.694 \pm 0.018$ & $4.57 \pm 0.22$   & $0.626 \pm 0.018$   & $0.3493 \pm 0.0074$ \\
      80 & 2.00     & 1.31     & $0.697 \pm 0.015$ & $4.54 \pm 0.19$   & $0.624 \pm 0.017$   & $0.3485 \pm 0.0066$ \\
      80 & 2.25     & 1.31     & $0.708 \pm 0.010$ & $4.41 \pm 0.13$   & $0.614 \pm 0.014$   & $0.3441 \pm 0.0051$ \\
      80 & 2.50     & 1.40     & $0.7241 \pm 0.0075$ & $4.22 \pm 0.10$ & $0.598 \pm 0.012$ & $0.3375 \pm 0.0042$ \\
      80 & 2.75     & 1.38    & $0.7288 \pm 0.0063$ & $4.167 \pm 0.087$ & $0.593 \pm 0.011$ & $0.3354 \pm 0.0038$ \\
      80 & 3.00     & 1.31    & $0.7296 \pm 0.0043$ & $4.157 \pm 0.066$ & $0.5922 \pm 0.0097$ & $0.3350 \pm 0.0031$ \\
       \hline\hline
       64 & 0.50    & 1.58     & 1   & $2.31  \pm 0.25$     & $0.452 \pm 0.059$    & $0.257\pm 0.028$ \\
       64 & 0.70    & 1.36     & 1   & $2.472  \pm 0.052$ & $0.494 \pm 0.015$    & $0.2745 \pm 0.0053$ \\
       64 & 0.80    & 1.29     & 1   & $2.469 \pm 0.049$  & $0.494 \pm 0.015$    & $0.2743 \pm 0.0049$ \\
      64 & 0.90     & 1.36    &  1   & $2.515 \pm 0.040$  & $0.506 \pm 0.012$     & $0.2789 \pm 0.0040$ \\
      64 & 1.00     & 1.29    &  1   & $2.521 \pm 0.028$  & $0.5082 \pm 0.0090$ & $0.2795 \pm 0.0027$ \\
      64 & 1.19     & 1.77    &  1   & $2.478 \pm 0.027$  & $0.4955 \pm 0.0089$ & $0.2752 \pm 0.0026$ \\
       \hline
      80 & 0.50      & 0.61    & 1   & $1.96  \pm 0.14$      & $0.378  \pm 0.033$ & $0.220 \pm 0.015$ \\
      80 & 0.70      & 1.15    & 1   & $2.531   \pm 0.064$ & $0.512 \pm 0.019$ & $0.2809 \pm 0.0064$ \\
      80 & 0.80      & 1.15    & 1   & $2.554   \pm 0.060$ & $0.519 \pm 0.018$ & $0.2832  \pm 0.0060$ \\
      80 & 0.90      & 1.15    & 1   & $2.578  \pm 0.056$ & $0.526 \pm 0.017$ & $0.2857 \pm 0.0055$ \\
      80 & 1.00      & 1.09    & 1   & $2.566  \pm 0.044$ & $0.522 \pm 0.014$ & $0.2845 \pm 0.0043$ \\
      80 & 1.10      & 1.14    & 1   & $2.525  \pm 0.036$ & $0.510 \pm 0.011$ & $0.2803  \pm 0.0034$ \\
      80 & 1.25      & 1.60    & 1   & $2.454  \pm 0.035$ & $0.489 \pm 0.011$ & $0.2733 \pm 0.0034$ \\
       \hline\hline
   \end{tabular}
   \caption{Fits to the Refined Gribov-Zwanziger functional form $D_{RGZ}(p^2)$.
   The upper part refers to the fits have $Z$ as a free parameter, while in the lower part of the table $Z = 1$.
    $\nu$ refers to the $\chi^2/d.o.f.$ For the smaller lattice (larger ensemble) we only show fits with a $\nu < 2$.
   All parameters are in powers of GeV.}
   \label{tab:rgzfits}
\end{table}

\section{Reproducing the Full Range of Momenta}

If the infrared lattice data is well described by the Refined Gribov-Zwanziger type of propagator, how can one extend this functional
form to cover the full range of lattice momenta, without compromising  the perturbative tail at high momenta? In order to achieve such goal,
we assume that the gluon propagator is given by
\begin{equation}
   D(p^2) = Z \, \frac{ p^2 + M^2_1}{p^4 + M^2_2 ~ p^2 + M^4_3} \Bigg[ \omega ~ \ln \left(\frac{p^2 + m^2_g(p^2)}{\Lambda^2_{QCD}}\right) + 1 \Bigg]^{\gamma_{gl}} \ ,
\end{equation}
where $\omega = 11 \, N \, \alpha ( \mu ) / 12 \, \pi$, $\alpha_s ( \mu )$ is the strong coupling constant at the renormalisation scale $\mu$,
$\gamma_{gl} = -13/22$ is the one-loop gluon anomalous dimension and $m^2_g(p^2)$ is a running mass that regularises the log function, such that
the Gribov-Zwanziger expression is recovered for momenta $p \lesssim 1$ GeV and at high momenta $D(p^2)$ reproduces the one-loop renormalisation
improved result. 

In our study we take the MOM-scheme where $\alpha_s (3 \mbox{ GeV}) = 0.3837$, see~\cite{Aguilar:2010gm}, and following the 
works~\cite{Boucaud:2008gn,Sternbeck:2010xu,Dudal:2017kxb} we set $\Lambda_{QCD} = 0.425$ GeV for the pure Yang-Mills theory. Note that in this
way the only fitted parameters are $Z$, $M^2_1$, $M^2_2$, $M^4_3$ and those that parametrise the regularisation mass $m^2_g(p^2)$.

\begin{figure}[t] 
   \centering
   \includegraphics[width=6in]{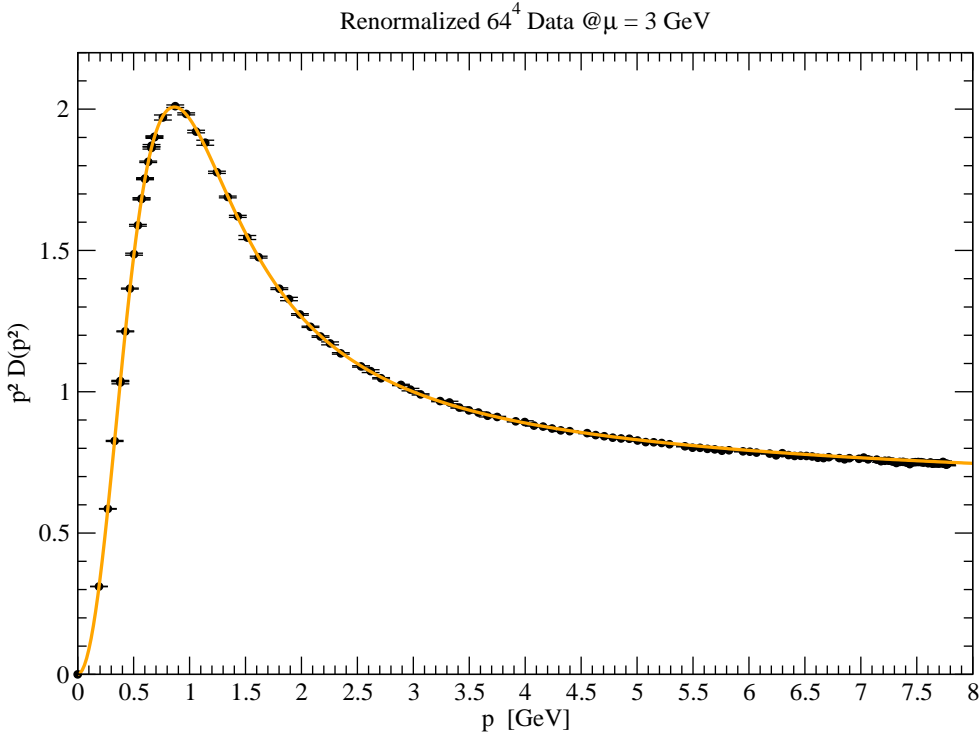} 
   \caption{Lattice data together with the global fit mentioned in the text.}
   \label{fig:example}
\end{figure}

We have tried several functional forms for the regularisation mass, see~\cite{Dudal:2018cli} for further details, and our best fit used
\begin{equation}
   m^2_g(p^2) = \lambda^2_0 + \frac{m^4_0}{p^2 + \lambda^2}
\end{equation}   
and resulted in $\chi^2/d.o.f. = 1.11$ with $Z =  1.36992(72)$, $M^2_1 = 2.333(42)$ GeV$^2$,
$M^2_2 = 0.514(24)$ GeV$^2$, $M^4_3 = 0.2123(32)$ GeV$^4$,
$m^4_0 = 1.33(13)$ GeV$^2$, $\lambda^2 = 0.100(35)$ GeV$^2$ and $\lambda^2_0 = −0.954(70)$ GeV$^2$ for the smaller lattice volume,
which had the ensemble with
the higher number of configurations. The lattice data together with the fit just mentioned can be seen on Fig.~\ref{fig:example}.
We call the reader attention to the good agreement between the values for $M^2_1$, $M^2_2$ and $M^4_3$ obtained in the global fit
and those reported in the previous section.

For the global fit discussed here, $D(p^2)$ predicts a pair of complex conjugate poles at $p^2 = - 0.257 \pm  i ~ 0.382$ GeV$^2$ and a pair of
complex conjugate branch points at momenta $p^2 = 0.43 \pm i ~ 1.02$ GeV$^2$. If the pair of complex conjugate poles are associated with momenta whose real part
is negative and the values of the poles are essentially independent of the regularisation mass $m^2_q(p^2)$, the computed branch points show a strong
dependence on the model used for $m^2_q(p^2)$.

\section{Results and Conclusions}

Our results show that the tree level propagator associated with Refined Gribov-Zwanziger action and the Very Refined Gribov-Zwanziger action
are compatible with the lattice data up to momenta $\sim 1$ GeV. In particular the results for the Very Refined Gribov-Zwanziger action translate into
a constraint on the condensates of the theory. From the point of view of the theory, it would be interesting to have high order predictions for the
Landau gauge gluon propagator to be tested against high precision simulations.

Our analysis also provide a global fit that results in a propagator with a pair of complex conjugate poles and a pair of complex conjugate branch points.
If the poles seem to be robust against the logarithmic regularisation mass, the same does not apply to the location of the branch points.

\section*{Acknowledgments}

The authors acknowledge the Laboratory for Advanced Computing at University of Coimbra for providing HPC computing resources Navigator that have 
contributed to the research results reported within this paper (URL http://www.lca.uc.pt). This work was granted access to the HPC resources of the 
PDC Center for High Performance Computing at the KTH Royal Institute of Technology, Sweden, made available within the Distributed European Computing Initiative by the
PRACE-2IP, receiving funding from the European Community's Seventh Framework Programme (FP7/2007-2013) under grand agreement no. RI-283493. The use of Lindgren 
has been provided under DECI-9 project COIMBRALATT. We acknowledge that the results of this research have been achieved using the PRACE-3IP project (FP7 RI312763)
resource Sisu based in Finland at CSC. The use of Sisu has been provided under DECI-12 project COIMBRALATT2. P. J. Silva acknowledges support by FCT under 
contracts SFRH/BPD/40998/2007 and SFRH/BPD/109971/2015. O. O. and P. J. S. acknowledge financial support from FCT Portugal under contract  
UID/FIS/04564/2016. Part of this work is included in INCT-FNA Proc. No. 464898/2014-5 project. O. O. acknowledges support from CAPES process 
88887.156738/2017-00 and FAPESP grant number 2017/01142-4. D. D. acknowledges financial support from KU Leuven IF project C14/16/067.


\end{document}